\documentclass[journal]{IEEEtran}

\newcommand{\etal}{{\it et~al.}}
\newcommand{\ie}{{\it i.e., }}

\usepackage{amsmath, scalerel}

\usepackage[linesnumbered,ruled]{algorithm2e}
\usepackage{graphicx}
\graphicspath{ {images/} }
\usepackage{pgfplots, lipsum}
\usepackage{textcomp}
\pgfplotsset{compat = newest}

\pgfplotsset{width=10cm,compat=1.9}
\usepgfplotslibrary{external}
\usepackage{tikz}
\usetikzlibrary{arrows}
\usepackage{caption}
\usepackage{siunitx}
\usepackage[font={small}]{caption}
\usepackage{amsmath}
\usepackage[export]{adjustbox}
\usepackage{lipsum}
\usepackage{subcaption}

\usepackage{xcolor}
\usepackage{pgfplots}
\usepackage{tikz}
\usepackage{pgfkeys}
\usepackage{url}
\usepackage{pgfplotstable}
\usepackage{subcaption}
\usepackage{filecontents}
\usepackage{multirow}
\usepackage{pgf}

\makeatletter
\def\BState{\State\hskip-\ALG@thistlm}

\makeatother

\usepackage{cite}

\ifCLASSINFOpdf
  
\else
  
\fi

\begin{document}

\title{ Improving Hierarchy Storage for Video Streaming in Cloud
}

\author{\IEEEauthorblockN{Mahmoud Darwich$^{1}$, Yasser Ismail$^{2}$}, Talal Darwich$^{3}$, Magdy Bayoumi$^{4}$\\ 
\IEEEauthorblockA{ $^{1}$Department of Mathematical and Digital Sciences, 
Bloomsburg University of Pennsylvania, PA 17815\\
$^{2}$Electrical Engineering Department, 
Southern University and A\&M College, Baton Rouge LA 70807\\
$^{3}$Microchip Technology Inc., San Jose, CA 95134\\
$^{4}$Department of Electrical and Computer Engineering,
University of Louisiana at Lafayette, LA 70504 \\
Email: mdarwich@bloomu.edu, yasser\_ismail@subr.edu, talal.darwich@microchip.com, magdy.bayoumi@louisiana.edu}
}

\maketitle

\begin{abstract}
Frequently accessed video streams are pre-transcoded into several formats to satisfy the characteristics of all display devices. Storing several video stream formats imposes a high cost on video stream providers using the old classical way. Alternatively, cloud providers offer a high flexibility of using their services and at a low cost relatively. Therefore, video stream companies adopted cloud technology to store their video streams. Generally, having all video streams stored in one type of cloud storage, the cost rises gradually. More importantly, the variation of the access pattern to frequently accessed video streams impacts negatively the storage cost and increases it significantly. To optimize storage usage and lower its cost, we propose a method that manages the cloud hierarchy storage. Particularly, we develop an algorithm that operates on parts of different videos that are frequently accessed and stores them in their suitable storage type cloud. Experiments came up with promising results on reducing the cost of using cloud storage by 18.75 \%.

\end{abstract}

\begin{IEEEkeywords}cloud, storage,
video stream, pre-transcoding, clustering
\end{IEEEkeywords}

\IEEEpeerreviewmaketitle

\section{Introduction }

The advances in display devices technology from big screens of TVs, computer monitors, etc., to small screens of mobile devices, allows people around the globe to afford to purchase these devices. In addition to social media applications, it opened the doors
to all users to watch a video stream and even stream their videos through their devices. Managing a huge number of videos became challenging for video stream providers. Gratefully, Cloud technology came up to offer a solution that helped video stream companies to process the video streams easily and less costly. However, the exponentially rapid growth of video streams number adds a further challenge in the could that needs to be addressed \cite{xiangbo2020}. 

Cloud services provide two essential operations on a video stream either transcoding or storing. Transcoding is an operation applied on a video stream to create another format, it requires extensive computational resources which are considered high cost, while cloud storing costs less because the computational virtual machines and storing servers  are charged in hourly  and monthly rates respectively \cite{amazon}.

 Several formats for each video stream are created to meet the characteristics of all devices specification\cite{netflix}. Accordingly, to avoid the high cost of transcoding, many formats of a video are pre-transcoded and stored in the cloud. The criteria for storing video formats is based on the frequent accesses to them, thus, They are called Frequently Accessed Video streams (FAVs).  
 
Amazon Web Services (AWS) is well known as a cloud services provider and is adopted in our research to implement the experiments. However, our approach could be applied to other cloud platforms.
Amazon storage offers a variety of storage options. In this research we consider the following:
\begin{itemize}
\item \texttt{S3 Standard}: It is typically used to store frequently accessed video streams.
\item \texttt{S3 Standard-Infrequent Access (S3 Standard-IA)}: is used to store infrequently accessed video streams.
\item \texttt{One Zone-Infrequent Access (S3 One Zone-IA)}: is used to store less infrequently accessed video streams.
\item \texttt{S3 Glacier}: stores frequently access video streams for long term.
\end{itemize}

The research problem in this work is how to store the part of the video which is frequently accessed in different cloud storages. Therefore, an approach that carries out clustering on the parts of different video streams is proposed to optimize the cloud storage cost.

The contribution of this research is summarized as follows: (1) We propose a method that decides parts of a video stream to be stored in different cloud storage. This is done through clustering of frequently accessed parts. (2) We analyze the efficiency of the proposed method when the number of accessed video streams varies.

Our proposed work differs from our previous work \cite{darwich2020} in that we propose a method that clusters the frequently accessed parts of video streams in different storage and thus it reduces the cloud cost.

The paper is organized as follows: Section II discusses the related works. In section III, the clustering method is detailed. Section IV reveals the experiment setup and section V presents the results. Section VI concludes the paper.

\section{Related Work}

Darwich \etal \cite{darwichhotness} proposed methods to manage frequently accessed video streams in a repository to reduce the cost of cloud services. They came up with an algorithm that measures the hotness of the frequently accessed video stream, their method stores the frequently accessed video streams. The results show a cost reduction of using cloud services.

Darwich \etal \cite{darwich2016} proposed an approach that decides whether a video stream should be stored in the cloud or re-transcoded upon request. Particularly, the calculated cost ratio of storing cost to transcoding cost. If the ratio is less than 1, the method will store the video stream. Their approach could reduce the cost significantly.

Darwich \etal \cite{darwich2020} proposed a method that stores frequently accessed videos in different types of cloud storage. They clustered frequently accessed video streams in the cloud storage which have low prices and thus their method could reduce the incurred cost.

Jokhio \etal \cite{jokhio} developed a method that reduces the cost of cloud services. Their method is based on calculating the popularity of video stream, they used a weighted graph to trade-off between transcoding storing operation. their proposed approach decides which video that should be stored or transcoded upon request.

Gao \etal \cite{gao} developed a scheme to manage the contents of video streams in media clouds. Particularly, they could split video streams into segments and then analyze them based on the view pattern. Their method decides which segments should be stored in the cloud and deletes the rest. The authors aim to minimize the operational cost of the cloud. Accordingly, They proposed an algorithm that could reduce the total cost by up to 30\%. 

Zhao \etal \cite{zhao} presented an approach to build a relationship between video views and its version. The authors came up with a strategy that uses a graph to form a relationship among video versions. Based on the graph, they could calculate the transcoding cost of each video in each version. Moreover, they compute the storage costs of videos. They analyze the transcoding and storage costs with views, their approach resulted in a reduction significantly in the cost of cloud services.

\section{Proposed GOPs Clustering Method}
\subsection{Entity of Video Stream}
\begin{figure}
 \includegraphics[width=7.47cm, height=10.87cm]{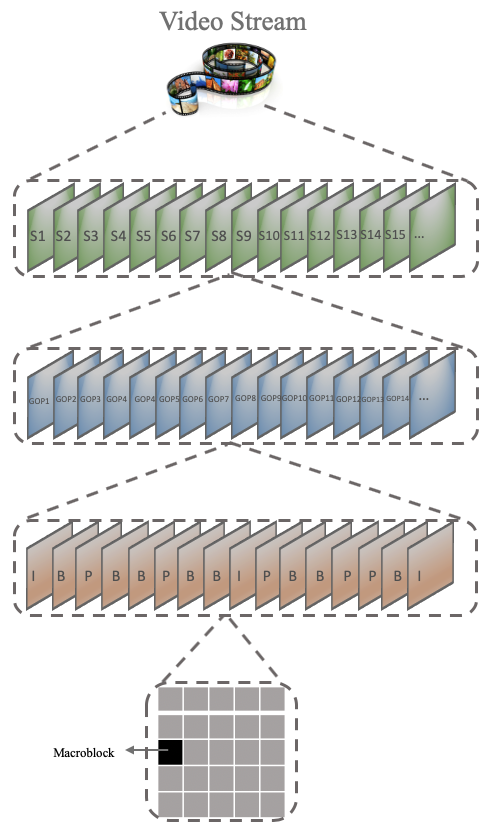}
 \caption{Structure of a video stream}
 \label {video-stream-structure}
\end{figure}

A video stream is built by many sequences as shown in Fig.\ref{video-stream-structure}. Each sequence is composed of Group Of Pictures (GOPs). Further, a GOP is made by different types of frames (\ie B (bi-directional) frames, P (predicted), and I (intra). Each frame is divided into small slices called macroblocks (MB) as shown in Fig.\ref{video-stream-structure}. Video processing is achieved at the GOPs level because they are processed independently \cite{jokhio1}. Thus in this research, we consider storing a video stream through storing its GOPs.

\subsection{Proposed Storing Scheme}
\begin{figure}
 \hfill\includegraphics[width=9cm]{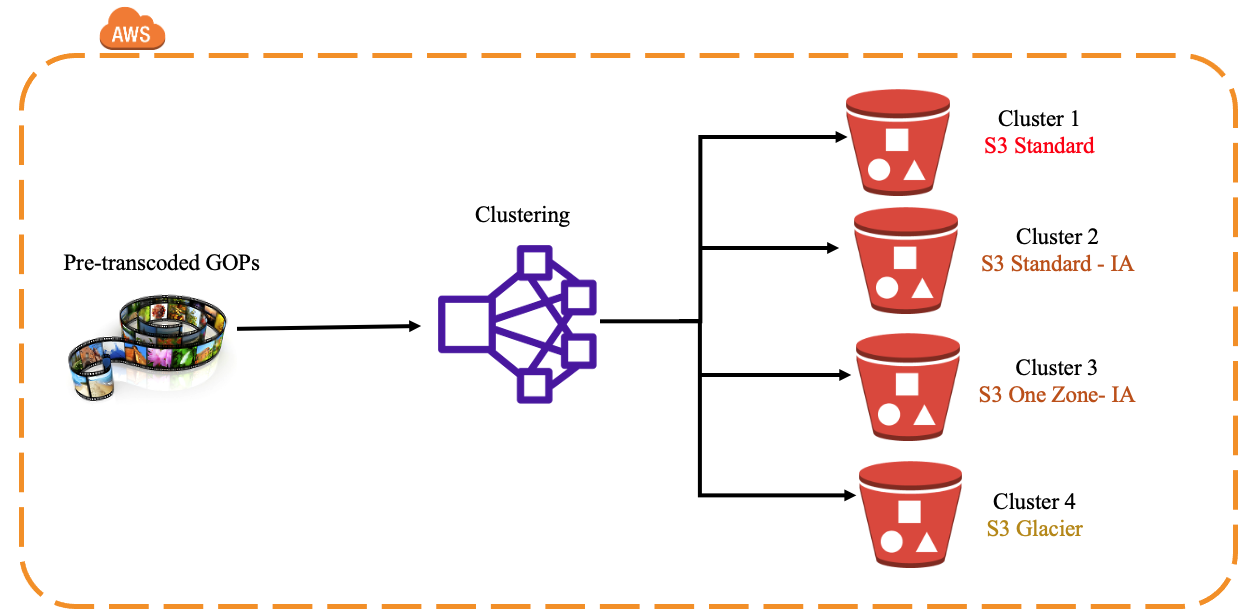}
 \caption{GOPs storing scheme in cloud}
 \label {storing_scheme}
\end{figure}
We present a scheme to manage media storage in the cloud in Fig.\ref{storing_scheme}. First, the scheme receives the GOPs that are decided to be stored (i.e., pre-transcoded), then these GOPs go through the clustering process. At this point, the proposed scheme analyzes each pre-transcoded GOP and distributes it to its suitable storage. The diagram provides four types of cloud storage; \texttt{S3 Standard}, \texttt{S3 Standard-IA}, \texttt{S3 One Zone-IA}, \texttt{S3 Glacier } are dedicated for cluster 1, cluster 2, cluster 3, and cluster 4 respectively.

\subsection{ Proposed algorithm }
The proposed algorithm is an enhanced version of the previous one \cite{darwich2020}. Its goal is to reduce the cost of storing frequently accessed video streams in the cloud. The algorithm mechanism applies clustering on the frequently accessed GOPs of video streams and then distribute them to their suitable storage in the cloud. The algorithm is executed on frequently accessed GOPs periodically and its pseudo-code is presented in algorithm \ref{al2}. The algorithm receives the size of each frequently GOP, its number of views, and the cloud storage price as input. The algorithm output clusters the frequently accessed GOPs into four clusters and then calculates their storage costs.

\begin{figure}
 \includegraphics[width=9cm, scale=1]{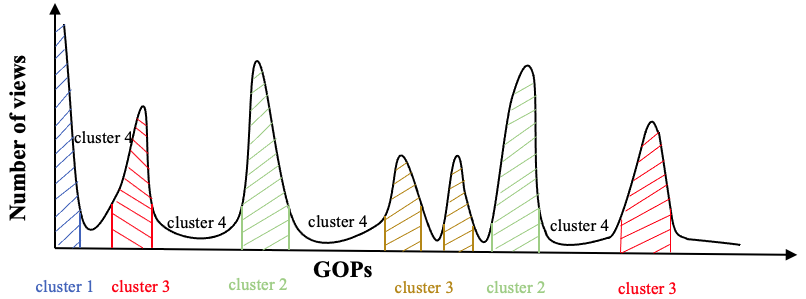}
 \caption{
 Clustering of frequently accessed GOPs in the long-tail distribution}
 \label {Video-partial4}
\end{figure}

 According to Miranda \etal \cite{miranda}, the access pattern to GOPs in a video stream follows the long-tail distribution. That means the beginning of video GOPs are watched frequently more than the remaining other GOPs. However, many videos could have random access patterns. In this case, many GOPs are frequently accessed by viewers. Those frequently accessed GOPs are distributed across the video starting at the beginning through the whole video as illustrated in Fig.\ref{Video-partial4}.

 In this case, the proposed algorithm clusters the frequently accessed GOPs into four clusters. The clustering process is based on a similar number of views of GOPs. Cluster 1, cluster 2, cluster 3, and cluster 4 contain GOPs that have a similar number of views (Step 1). Then, the storage cost is computed for each cluster (Step 2 - Step 5). In Step 6, the total storage cost of all clusters is calculated.

{\SetAlgoNoLine%

\begin{algorithm}

 \SetKwInOut{Input}{Input}
 \SetKwInOut{Output}{Output}

 \Input{Frequently accessed $GOPs $: $GOP_{i}$\\
 	 Size of each frequently accessed $S_{GOP_{i}} $ \\
	 Cloud Storages price: $P_{S_{1}}$, $P_{S_{2}}$, $P_{S_{3}}$, $P_{S_{4}}$\\
	 Number of views of each frequently accessed $ GOP_{i}$\\

 }
 \Output{Storage Cost of all frequently accessed $GOPs$\\ }
 
 Apply K-Means clustering on all frequently accessed $ GOPs$ with $K=4$\\
 Storage Cost of GOPs Cluster 1: $C_{S_{1i}}\leftarrow\dfrac{\sum S_{GOP_{j}} \cdot P_{S_{1}}}{2^{10}}$\\
 Storage Cost of GOPs Cluster 2: $C_{S_{2i}}\leftarrow\dfrac{\sum S_{GOP_{j}} \cdot P_{S_{2}}}{2^{10}}$\\
 Storage Cost of GOPs Cluster 3: $C_{S_{3i}}\leftarrow\dfrac{\sum S_{GOP_{j}} \cdot P_{S_{3}}}{2^{10}}$\\
 Storage Cost of GOPs Cluster 4: $C_{S_{4i}}\leftarrow\dfrac{\sum S_{GOP_{j}} \cdot P_{S_{4}}}{2^{10}}$\\
 Total storage cost of frequently accessed $ GOPs$: $C_{S_{GOPs}}\leftarrow C_{S_{1i}} + C_{S_{2i}}+ C_{S_{3i}} + C_{S_{4i}} $\\

 \caption{GOPs storing cost}
 \label{al2}
  
\end{algorithm}
}

\section{Experiment Setup }

\subsection{Videos Synthesis}
The experiments are implemented using synthesized video streams. We created repositories of video streams by applying the method in \cite{darwich2016} because we did not have an access to repositories of video stream providers.

\subsection{ Amazon Storage Rates}

Amazon Website Service (AWS) offers to users several types of storage for different purposes of data storing. Table \ref{tab:storageprice} shows the rates in USD.
 \begin{table}[ht]
\centering 
\caption{Amazon storage types and their rates in USD }
\begin{tabular}{c c c c} 
\hline\hline 
 Storage & Price \\ [0.5ex] 
\hline 

\texttt{S3 Standard } & \$0.023 GB/month \\
\texttt{S3 Standard-IA } & \$0.0125 GB/month \\
\texttt{S3 One Zone-IA } & \$0.01 GB/month\\
\texttt{S3 Glacier } & \$0.001 GB/month\\

\hline
\hline 
\end{tabular}

\label{tab:storageprice}
\end{table}

\subsection{ Methods for Comparison}

We assess the proposed method by comparing it to previous works. The description of previous works are summarized as follow:
\begin {itemize}
\item \emph{Fully pre-transcoding} method: it stores the whole video streams
\item \emph{Fully re-transcoding} method, it deletes the video stream and transcodes it upon request
\item \emph{Partial pre-transcoding} method in \cite{darwich2016} , it stores a part of video stream which receives frequent accesses in the cloud standard storage \texttt{S3 Standard}.
\item \emph{Clustering video streams} method in \cite{darwich2020}. It clusters the frequently accessed video stream and stores them in different storages in the cloud

\end{itemize}

\section { Simulation Results}


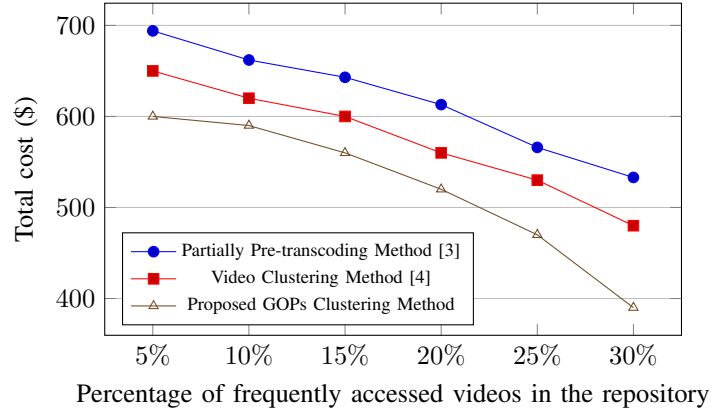
\begin{figure}
\centering
\begin{tikzpicture} 
 \begin{axis}[
 width =0.51*\textwidth,
 height = 6cm,
 ymajorgrids=true,
 legend style={font=\fontsize{7}{5}\selectfont},
 ylabel = {Total cost (\$)},
 xlabel={Percentage of frequently accessed videos in the repository},
 xticklabel={$\pgfmathprintnumber{\tick}\%$},
 xtick = data,
 legend pos= south west,
 ]
 \addplot
 coordinates {(5, 694)
 (10, 662)
 (15, 643)
 (20, 613)
 (25, 566)
 (30, 533)};

 \addplot
 coordinates {(5, 650)
 (10, 620)
 (15, 600)
 (20, 560)
 (25, 530)
 (30, 480)};

\addplot+[mark=triangle]
 coordinates {(5, 600)
 (10, 590)
 (15, 560)
 (20, 520)
 (25, 470)
 (30, 390)};
  
 \legend{ Partially Pre-transcoding Method \cite{darwich2016}, Video Clustering Method\cite{darwich2020}, Proposed GOPs Clustering Method}

 \end{axis}
  
\end{tikzpicture}


\caption{Cost comparison of the three methods, partially pre-transcoding method, video clustering method, and proposed GOPs clustering method when the number of frequently accessed videos varies }
 \label {favs}
\end{figure}

 \begin{table*}[ht]
\centering 
\caption{ Cost comparison in USD when the percentage of frequently accessed video changes }
\begin{tabular}{c c c c c c } 
\hline\hline 
 FAVs \% & Fully Re-Transconding & Full Pre-Transcoding& Partial Pre-Transcoding \cite{darwich2016}& Videos Clustering \cite{darwich2020} & Proposed GOPs Clustering \\ [0.5ex] 
\hline 

5\% &1596& 839 & 694 & 	 650 & 600		 \\
10 \% &1596& 	842	&662 &	620 &	590 	\\
15 \% &1596&	863	& 643 &	600 &	560 	\\
20 \% &1596&	947	&613 &	560 &	520	\\
25 \% &1596& 1424 & 566 & 530 & 470  \\
30 \% &1596& 3137& 533 &  480 & 390\\
\hline
\hline 
\end{tabular}
\label{tab:storageprice}
\end{table*}
 Table \ref{tab:storageprice} displays the results of the methods. The fully re-transcoding has the highest cost because each time the video stream is accessed, the video streams provider pays for using the virtual machines to transcode all video streams. 
 
 The fully pre-transcoding method has the second-highest cost because the video stream provider pays for using the storing servers to store all videos. It is worth mention that the cost of renting the virtual machines to transcode videos is higher than the cost of storing servers in the cloud.

 The partial pre-transcoding has a better performance compared to the two previously discussed methods. As the percentage of the frequently accessed video streams increases, this method works better because more parts of video streams are stored and reduce the cost further.
 
 The videos clustering method outperforms the previous method because it stores the frequently accessed video streams in different storages in the cloud, which makes the cost to be less than the cost of storing all video streams in one standard storage S3. As the percentage of frequently accessed video streams increase the cost is reduced further.
 
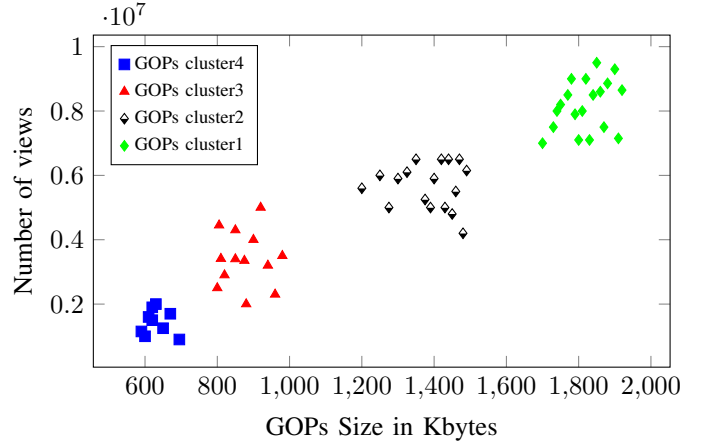
\begin{figure}
\begin{tikzpicture}

	\begin{axis}[%
	scatter/classes={%
		a={mark=square*,blue},%
		b={mark=triangle*,red},%
		c={mark=halfdiamond*},%
		d={mark=diamond*,green}},
		legend style={font=\fontsize{7}{5}\selectfont},
 ylabel = {Number of views},
 xlabel={GOPs Size in Kbytes},
 legend pos= north west,
 width =0.51*\textwidth,
 height = 6cm,
]
	\addplot[scatter,only marks,%
		scatter src=explicit symbolic]%
	table[meta=label] {
x y label
600 1000000 a 
620 1500000 a
630 2000000 a 
650 1250000 a
670 1700000 a 
695 900000 a
610 1600000 a 
620 1900000 a
590 1150000 a

800 2500000 b
850 3400000 b
805 4450000 b
810 3410000 b
820 2900000 b
850 4300000 b
875 3350000 b
880 2000000 b
900 4000000 b
920 5000000 b
940 3200000 b
960 2300000 b
980 3500000 b

1250 6000000 c
1275 5000000 c
1300 5900000 c
1325 6100000 c
1350 6500000 c
1375 5250000 c
1390 5000000 c
1200 5600000 c
1400 5900000 c
1420 6500000 c
1430 5000000 c
1440 6500000 c
1450 4800000 c
1460 5500000 c
1470 6500000 c
1480 4200000 c
1490 6150000 c

1700 7000000 d
1730 7500000 d
1740 8000000 d
1750 8200000 d
1770 8500000 d
1780 9000000 d
1790 7900000 d
1800 7100000 d
1810 8000000 d
1820 9000000 d
1830 7100000 d
1840 8500000 d
1850 9500000 d
1860 8600000 d
1870 7500000 d
1880 8860000 d
1900 9300000 d
1910 7150000 d
1920 8650000 d
	};
\legend{ GOPs cluster4, GOPs cluster3,GOPs cluster2,GOPs cluster1}
 
	\end{axis}
	
	\label{cluster}
	
\end{tikzpicture}
\caption{Clustering frequently accessed GOPs based on the number of views}
\label{gop_clusters}
\end{figure}

 The proposed method overcomes the performance of all previously discussed methods. Frequently accessed GOPs are clustered and stored in different cloud storages as shown in Fig. \ref{gop_clusters}. Cluster 1 contains the frequently accessed GOPs that have the highest views. Those GOPs are more demanded by viewers and thus stored in \texttt{ S3 Standard-IA}, which has the fastest access and highest cost among the others. cluster 2 contains all GOPs that have similar views. Its GOPs are less accessed than cluster 1, they are stored in \texttt{S3 One Zone-IA} which costs less than \texttt{ S3 Standard-IA}. The GOPs, which are less demanded, are included in Cluster 3 and cluster 4 which have low costs. When the percentage of frequently accessed GOPs increases the proposed method reduces the incurred cost by 18.75\% compared to the videos clustering method and 27\% compared to the partial pre-transcoding method.
 
 For the sake of clarity, Fig. \ref{favs} presents the comparison of partially pre-transcoding, video clustering, and proposed GOPs clustering methods, we eliminate the first two methods because of their high numbers which deteriorate the curves visibility.

\section{Conclusion}
In this paper, our proposed method reduced the cost of the cloud services to store video streams. Particularly, our method clusters the frequently accessed GOPs and store them in four different storages, cluster 1 contains GOPs that has the highest views, which provides fast access to it, while clusters 2, 3, and 4 are for the GOPs with fewer views. The experiment results show that the proposed method reduces the cost by 18.75\% compared to the video clustering method and 27\% compared to the partial pre-transcoding method.


\begin{thebibliography}{1}
\bibitem{xiangbo2020}
Xiangbo Li, Mahmoud Darwich, Mohsen Amini Salehi, Magdy Bayoumi "A survey on cloud-based video streaming services". Advances in Computers, Elsevier, 2021, ISSN 0065-2458

\bibitem{netflix}

http://techblog.netflix.com/2012/12/videos-of-netflix-talks-at-aws-reinvent.html
\bibitem{darwich2016}
Darwich, Mahmoud, Ege Beyazit, Mohsen Amini Salehi, and Magdy Bayoumi. "Cost efficient repository management for cloud-based on-demand video streaming." In 2017 5th IEEE International Conference on Mobile Cloud Computing, Services, and Engineering (MobileCloud), pp. 39-44. 
\bibitem{darwich2020}
Darwich, Mahmoud, Yasser Ismail, Talal Darwich, and Magdy Bayoumi. "Cost-Efficient Storage for On-Demand Video Streaming on Cloud." In 2020 IEEE 6th World Forum on Internet of Things (WF-IoT), pp. 1-4. 
\bibitem{amazon}
https://aws.amazon.com/s3/pricing/. Accessed December 2020.
\bibitem{jokhio}
F. Jokhio, A. Ashraf, S. Lafond, and J. Lilius, \emph{``A Computation and Storage Trade-off Strategy for Cost-Efficient Video Transcoding in the Cloud,''} 39th Euromicro Conference on Software Engineering and Advanced Applications, 2013.

\bibitem{zhao}
Zhao, Hui, \etal \emph{``A version-aware computation and storage trade-off strategy for multi-version VOD systems in the cloud.''} 20th IEEE Symposium on Computers and Communication (ISCC), 2015.


\bibitem{miranda}
Miranda, Lucas CO, Rodrygo LT Santos, and Alberto HF Laender. \emph{``Characterizing video access s in mainstream media portals.'' } Proceedings of the 22nd International Conference on World Wide Web, 2013.

\bibitem{darwichhotness}
Darwich, Mahmoud, Mohsen Amini Salehi, Ege Beyazit, and Magdy Bayoumi. "Cost-efficient cloud-based video streaming through measuring hotness." The Computer Journal 62, no. 5 (2019): 641-656.



\bibitem{gao}
G. Gao, W. Zhang, Y. Wen, Z. Wang and W. Zhu,\emph{``Towards Cost-Efficient Video Transcoding in Media Cloud: Insights Learned From User Viewings,''} in IEEE Transactions on Multimedia, vol. 17, no. 8, pp. 1286-1296, Aug. 2015.

\bibitem{jokhio1}
Jokhio, Fareed, \etal \emph{``Analysis of video segmentation for spatial resolution reduction video transcoding.''} International Symposium on Intelligent Signal Processing and Communications Systems (ISPACS), 2011.





 

















\end{thebibliography}
\end{document}